\documentclass[sigconf]{acmart}

\newcommand{\etal}{{\textit{et al.} }}
\newcommand{\ie}{{\textit{i}.\textit{e}. }}

\newcommand{\etc}{\textit{etc}.}
\newcommand{\ignore}[1]{}
\newcommand{\zztitle}[1]{\vspace{5pt}\noindent\textbf{#1.}}

\usepackage{makecell}
\usepackage{multirow}
\usepackage{tablefootnote}
\usepackage{caption}
\usepackage{subcaption}

\makeatletter
\g@addto@macro{\UrlBreaks}{\UrlOrds}
\makeatother

\setcopyright{none} 
\acmConference[Anonymous Submission to ACM CCS 2019]{ACM Conference on Computer and Communications Security}{Due 19 May 2017}{Dallas, Texas}
\acmYear{2017}

\begin{document}
\title{Do Not Return Similarity: Face Recovery with Distance} 

\begin{abstract}
Machine Learning (ML) has already been integrated into all kinds of systems, helping developers to solve problems with even higher accuracy than human beings. However, when implementing a system with ML model embedded, developers may accidentally take not enough care of the outputs of ML models, resulting in severe consequences like hurting data owners' privacy. Specifically for face authentication systems, we show that profile photos can be recovered once the system shows similarity (distance) to users.

In this work, we focus on understanding the risks of leaking distances of faces calculated by face recognition models. To show the consequence, we reveal several kinds of scenarios in which distances are accidentally leaked. E.g., a face verification system deployed by a government organization leaks similarity to attackers. Further, as we firstly identified, with leaked distances, attackers can easily recover the victim's face embedding in the authentication database, when using our proposed method. Threaten is that the profile photo corresponding to the embedding can then be recovered, indicating devastating consequences to authentication system security and users' privacy. This is achieved with our devised GAN-like model, which showed 93.65\% success rate on popular face embedding model.

\end{abstract}

\author{Mingtian Tan, Zhe Zhou\\Fudan University\\\{18210240176, zhouzhe\}@fudan.edu.cn}


\maketitle

\section{Introduction}
Machine learning (ML) gained a great success because of its high accuracy and convenience to implement. ML outperforms both human beings and traditional logic based programs for a lot of problems, like face recognition, board game \etc, which convinces people to confidently integrate ML models to various applications. Apart from high accuracy, ML is easy to implement, which also contributed to its popularity. Usually a deep learning model costs developers several hundreds of Python codes but can already produce satisfied accuracy.

Face recognition (FR) benefit a lot from ML and are employed by numerous real world systems, but FR techniques are also blamed for privacy issues. With recent deep learning techniques involved, face embedding, the supporting technique behind FR, harvested a big step forward in terms of accuracy, which realized applications like face authentication, face search \etc. As these applications became popular, huge amount of face photos are processed by FR. With no doubt, these photos are highly sensitive and should be never exposed to unauthorized people. Therefore, researchers started their work around privacy leakages resulted from FR. To avoid potential privacy invasion, San Francisco even banned the use of FR techniques. However, the privacy concerns didn't impede the popularity of FR and related applications.

FR techniques may unintentionally leak users' face photo when deployed in face authentication systems, as we firstly identified. To show the severeness, in this work, we demonstrate a kind of attack that targets face authentication systems. The attack allows an attacker to pass authentication and even steal profile photos of all users of a face authentication system, although actually authentication systems do not store any photo for users. The attack is not a result of bugs of a specific family of ML models but stems from leakages of face comparisons. The attack bases on three insights we identified in face recognition powered authentication systems: 1) Even a failed authentication result leaks information to attackers. 2) The information can be accumulated so that the face embedding, internal representation of a user, can be recovered. 3) The face embedding is as equal sensitive as the original face photo, because an attacker can easily recover the photo when she acquires an embedding.

\zztitle{Information leakage during face authentication} Every time face authentication is launched, the system leaks a small portion information about the claimed user's face to the prover via the comparison results, which could help attackers to recover the embedding (a vector representing a face) of the profile photo once enough information is collected. Specifically, the leakage results from the similarity attached in verification results. As we noticed, when implementing face authentication systems, developers sometimes show the similarity from the prover to the claimed user besides telling if face matched. The similarity actually leaks a small portion of information about the claimed user no matter if the prover is indeed the claimed user.

\zztitle{Embedding recovery with information leakage} At the first glance, exposing the similarity to users would not result in profile photo leakage, as the amount of information provided by the similarity is negligible comparing with the amount of information a profile photo. However, the information in different rounds of authentication can be accumulated, when the attacker actively try to launch authentication with different provers. When enough information is collected, as we will later show, an attacker can readily recover the embedding of the victim's profile photo, mainly because embeddings are actually high-dimension vectors which still obey algebraic geometry theorems. Therefore each authentication result signals the attacker an equation about the embedding. When enough amount of equations about the embedding are acquired, the attacker can work the embedding out.

\zztitle{Photo recovery with embedding} What's worse is that with the embedding of a victim's photo, the attacker can recover the profile photo with a GAN-like recovery model proposed by us. The embedding of a profile photo is generated by a deep learning model which is deemed to be super non-linear and hard to interpret. Therefore, to the best of our knowledge, there is no tool or algorithm to help attackers to get the reverse mapping of a deep learning model. Besides, the embedding process is generally a lossy function from higher input dimension to lower output dimension, so the reversing is theoretically difficult because of the lost information during mapping. As a result, there is no straightforward method to reverse a face embedding to its original profile photo.

After analyzing the face embedding model structure, we propose a deep learning model that maps face embeddings back to profile photos. The model tries to get back information lost during embedding with the help of GAN. The recovery model has a GAN like structure and produces satisfied recovery quality. Unlike GAN generators that map noises to data with given distribution, our recovery model, though also has a generator, tries to get a reverse mapping of a face embedding model, which is more challenging than training a normal GAN, because We need the model generates not only authentic face images but also face images corresponding to the embedding the attacker recovered. To force the GAN generator to generate embedding-face corresponded mapping, we devise a series of loss functions and special training process. As a result, the recovery quality is very high. An evaluation metric is that 93.75\% of the generated photos are recognized as from the embedding owner by the face embedding model.

The consequence of the face recovery is devastating. Take a face authentication based door entrance system of a building as an example, an attacker can claim to be arbitrary one of the legitimate tenant (victim) and try to pass the entrance with the recovered victim face. More severely, the method eventually can help attackers dump the whole profile photo database of the system. Theoretically, the attacker can enumerate all tenant ID and repeat the attack. Threatening is that the attacker knows nothing about the victim in this case.


To avoid any further embedding leakages, we suggest that ML library developers must clearly state in the documents that the embedding is as equal sensitive as original image. Also, any distance or similarity calculated with embeddings should not be freely exposed to users or third parties. Beside, we suggest ML researchers design embedding models that cannot be easily reversed, \ie one-way ML model.

We summarize our contributions as follow:
\begin{itemize}
    \item We identified that face authentication developers may leak a small portion information about the claimed user when showing distances (similarities) from the prover to the claimed user.

    \item We show that such information can be collected to recover the embedding of the claimed user. To show the feasibility, we construct an equation set to help attackers work out the claimed user's embedding with query embeddings and leaked distances.

    \item We discovered that an embedding is as equal sensitive as the behind profile photo.

    \item We designed a GAN-like recovery model to map face embeddings back to profile photos. With the model we showed that the recovery model can be applied to all the popular FR models we evaluated.
\end{itemize}

\section{Background}
We introduce some background techniques that will be used in this work.
\subsection{Face Embedding}


Face embedding is simply a mapping ($y=f(x)$) from the image ($x$) space to the vector space ($y$). The mapping could be useless unless important properties on $x$ also holds on $y$. Usually, people expect the similarity of embeddings is proportional to similarity of faces, so that the similarity of a pair of faces can be directly calculated by calculating the similarity of their embeddings (1- their distance).

With face embedding, a lot of tasks related to faces can be easy to implement. For example, face searching becomes to similar vector searching problem for which there are already tons of algorithms, like Locality Sensitive Hashing (LSH~\cite{gionis1999similarity}). Another example could be face verification: tell if two faces are from the same person. With face embedding, developers need only check if the distance between the two embeddings exceeds a pre-defined threshold.

\ignore{
\zztitle{Training face embedding models} A face embedding model is usually trained like shown by Figure.~\ref{fig:embedding}. Training profile photos are sent to an embedding model which usually is a deep convolution neural network. The network outputs fixed length vectors that will later be the embeddings. The vectors are connected to a simple classifier whose output error penalizes the whole model. 

\begin{figure}
    \centering
    \includegraphics[width=0.5\textwidth]{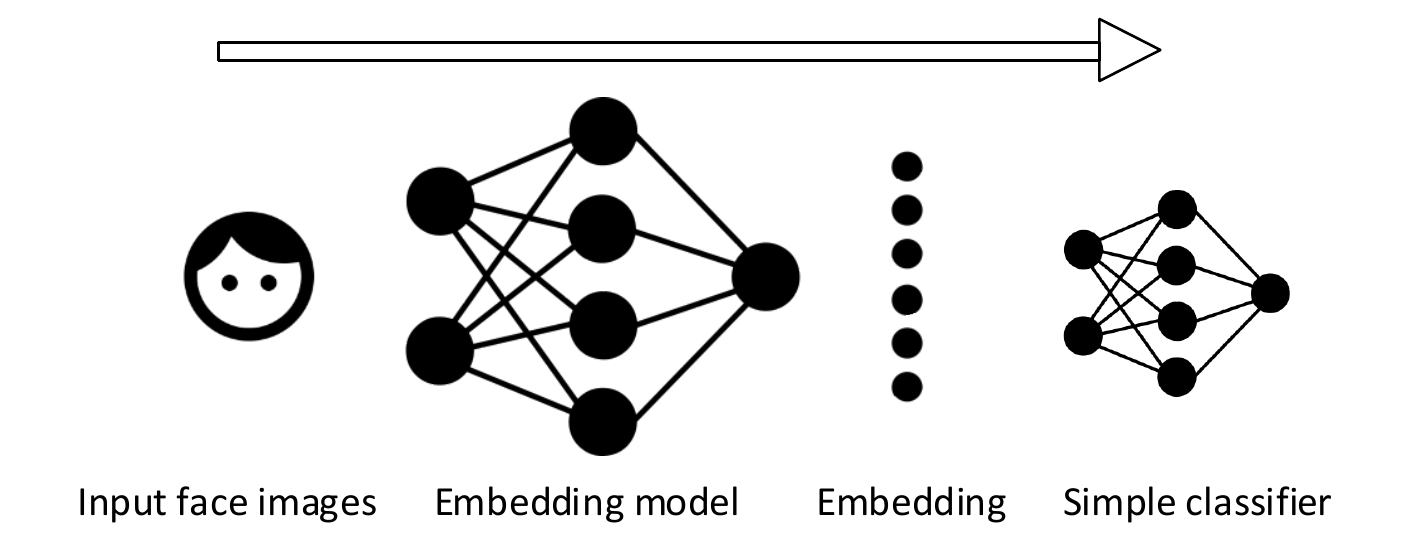}
    \caption{Training a face embedding model}
    \label{fig:embedding}
\end{figure}

With this training method, the distance of a pair of embeddings can represent the distance of their original profile photos because the classifier is nearly as simple as finding the class that has the smallest distance from the embedding. Specifically, a common implementation of such a classifier is just a fully connected layer followed by a softmax layer. The fully connected layer ($w \cdot x+b$) simply calculates the dot product (angle cosine) from the embedding against each class center. Then the softmax normalize them to probabilities of belonging to each class. }

\subsection{GAN}

Generative Adversarial Network (GAN) \cite{goodfellow2014generative} aims at synthesizing data of similar distribution to training data. Usually, GANs are used to generate photos that look authentic to human beings.

\zztitle{GAN structure} GANs usually come with two parts: a generator and a discriminator. The generator maps noises to the target space, while the discriminator determines if a given piece of data is authentic or synthesized (generated). During training, the generator is encouraged to fool the discriminator, \ie improve its generation capability, while the discriminator is forced to tell authentic and synthesized data apart, \ie improve its discrimination capability. Starting from random parameters, the generator has no ability to generate authentic data and the discriminator neither has ability to tell apart synthesized data. They are trained in turn to defeat each other, \ie have stronger capability than the other. With training process goes on, the two capabilities gradually increase in turn. As a result, the generator has strong enough generation capability to even fool human beings.

\zztitle{Conditional GAN (cGAN)~\cite{gauthier2014conditional}} Early GANs had no control over the generated data. However people wanted to generate data not only falling to a specific distribution but also with given attributes. To satisfy these demands, cGAN was proposed. In cGANs, the generator takes as input not only noises but also a given label, indicating generating data falling to the distribution of a specific class. Similarly, the discriminator also gets the label and tells if the input data are in the distribution and at the same time go with the class. As a result, the generator can generate data of the given label. 

There are also a bunch of famous GAN variant works, like image to image translation~\cite{isola2017image}, image to image translation without paired data~\cite{zhu2017unpaired}.

\subsection{PCA and SVD}
Principal component analysis (PCA) is a statistical tool to transform possibly correlated data into linearly uncorrelated variables (principal components). With the decomposed uncorrelated components, analysts can check the variance (importance) of each components and then try to compress the data by omitting those components with only little variance.  
Singular Value Decomposition (SVD) is a popular method to do PCA over matrices. In SVD, a $m$-by-$n$ matrix $M$ where each row is a piece of $n$ dimension observation and totally $m$ observations can be decomposed to the product of three matrices, \ie $M=U\cdot\Sigma \cdot V^\intercal$, in which $U$ and $V^\intercal$ are unitary matrices and $\Sigma$ being a rectangular diagonal matrix. When $M$ is not a square matrix, say, when $m > n$, $U$ will be $m$-by-$n$, $\Sigma$ being $n$-by-$n$ and $V$ being $n$-by-$n$. Each row of $V$ here is a component and the value corresponded in the diagonal of $\Sigma$ represents the variance of the component. Because $U$ and $V^\intercal$ are unitary, each column of $U$ or each row of $V$  will be orthogonal, indicating orthogonal components. People usually makes the diagonal of $\sigma$ in descending order.

When doing PCA, analysts usually want to approximate $M$ with another matrix $\tilde{M}$ that has a lower rank, which can be done with SVD by truncating $U$, $\Sigma$ and $V$. To have a rank $r$ $\tilde{M}$, we can truncate the matrices into $m$-by-$r$, $r$-by-$r$ and $n$-by-$r$, from top to down and from left to right. Because $\Sigma$ has descending order diagonal, the truncated dimensions have less variance. As a result, a $n$ dimensional row vector in $M$ can actually be represented as a $r$ dimensional vector with acceptable errors. 
\section{Overview \& Adversary Model}
We firstly briefly introduce the attack model and then discuss the capabilities we assume for the adversaries.

\subsection{Attack Overview}
The attacker aims at acquiring a profile photo of the target victim who is a user of a face authentication system. The attacker is interested in the profile photo because 1) She is interested in the privacy of the victim or 2) She wants to use the photo to pass the face authentication system on the behalf of the victim.

To achieve this, the attacker comes to the face authentication system that the victim already enrolled. She claims to be the victim by entering the ID of victim and then face authentication is started. She presents the system either a profile photo or a real person, depending on whether the system comes with a liveness detection system, but anyway obviously she cannot pass the authentication and the authentication system return to the attacker an authentication failure notice as well as the similarity that is below the threshold. The attacker records the photo causing the failure and the returned similarity, so she gets a <photo, similarity> pair. The attacker repeats this process several times with different provers, \ie different photos or different people. As a result, she gets a list of <photo, similarity> pairs.

With the list of <photo, similarity> pairs, the attacker can then 1) firstly recover the embedding of the victim's face according to our embedding recovery method, and then send the embedding to the GAN-like recovery model for image recovery.

\subsection{Embedding Model Assumptions}
Our later introduced attack method needs the involvement of embedding model inside the authentication system. In this section, we introduce our different assumptions about the embedding model the attacker can acquire. Our later method works with these assumptions differently.

\zztitle{White Box adversaries}
Most of times, the attacker knows the target model in detail, including structure, hyper-parameters and weights. In this scenarios, she is expected to recover images with better quality. We call this kind of adversary white box adversary.

An attacker may have multiple ways to acquire the model. For example, she can purchase or download for free the same face authentication software as the target system, and then reverse engineer the software to extract the model out. Reverse engineering indeed costs the attacker much effort but she needs only solve it once for all. According to our discussion with face recognition system engineers, over 80\% main stream face recognition solutions come from the top 10 CV companies, including SenseTime and Megvii \etc~In this case, attackers can identify which model is used and then get exactly the same model by purchasing from these AI solution companies. Besides, it would be easy if the system comes with open source face recognition library like Open face~\cite{amos2016openface}. In fact open source solutions already provide enough precision. If the system uses open source library, the attacker can then directly download and extract the model out after having identified which library is used.

\zztitle{Black Box Adversaries}
Some adversaries do not know the model in detail but can querying the model, in which case we call them black box adversaries.

Sometimes, extracting the model out from the software could be difficult, but exposing the interface for specific functions could be easy. If an adversary only gets the interface for embedding but could not get the model, she becomes to be a black box adversary. For some face authentication systems with network, they do not come with an embedding model inside. They use online model instead, mainly due to intelligence property protection to the model, which makes directly extracting the model out impossible. But in this case, the attacker still can use the model to acquire embeddings for her photos, if only she could make clear the API of the online embedding service, which requires much less reversing engineering work, comparing with the white box case. Besides, we found some systems employ public online face embedding service providers like Clarifai. They can be easily identified by the IP addresses of their servers. In this case, attackers can directly use the online model as a black box. 

A black box adversary can bootstrap to a white box adversary, by training a substitutional model. As we noticed that some model recovery method requires only query to a model but can work out a model that has a similar functionality as the imitated one. The most straightforward method is model distillation~\cite{hinton2015distilling}, which was originally used to compress a large model to be a smaller one, but was used as a tool to learn AI models in a lot of black box scenarios.

\zztitle{No Box Adversaries}
In some cases, the adversary has no any access to the model, or even the system itself. In this case, the adversary has no way to map the <photo, distance> list to <embedding, distance> list, which may causes troubles for the attack. However, as we will later show, the embedding spaces of different embedding models have parallel relationship in some extent, which means attackers can work out an embedding in her designated embedding space by using her own model instead of the one used to calculate the distance. Therefore, attackers can still recover the victim image using our method.

\section{Embedding Recovery}
\label{sec:emb}
In this section, we show that how embeddings can be worked out by attackers with only distance information leaked by the authentication system.

\begin{figure*}
\begin{subfigure}{0.65\textwidth}
    \centering
    \includegraphics[width=\textwidth]{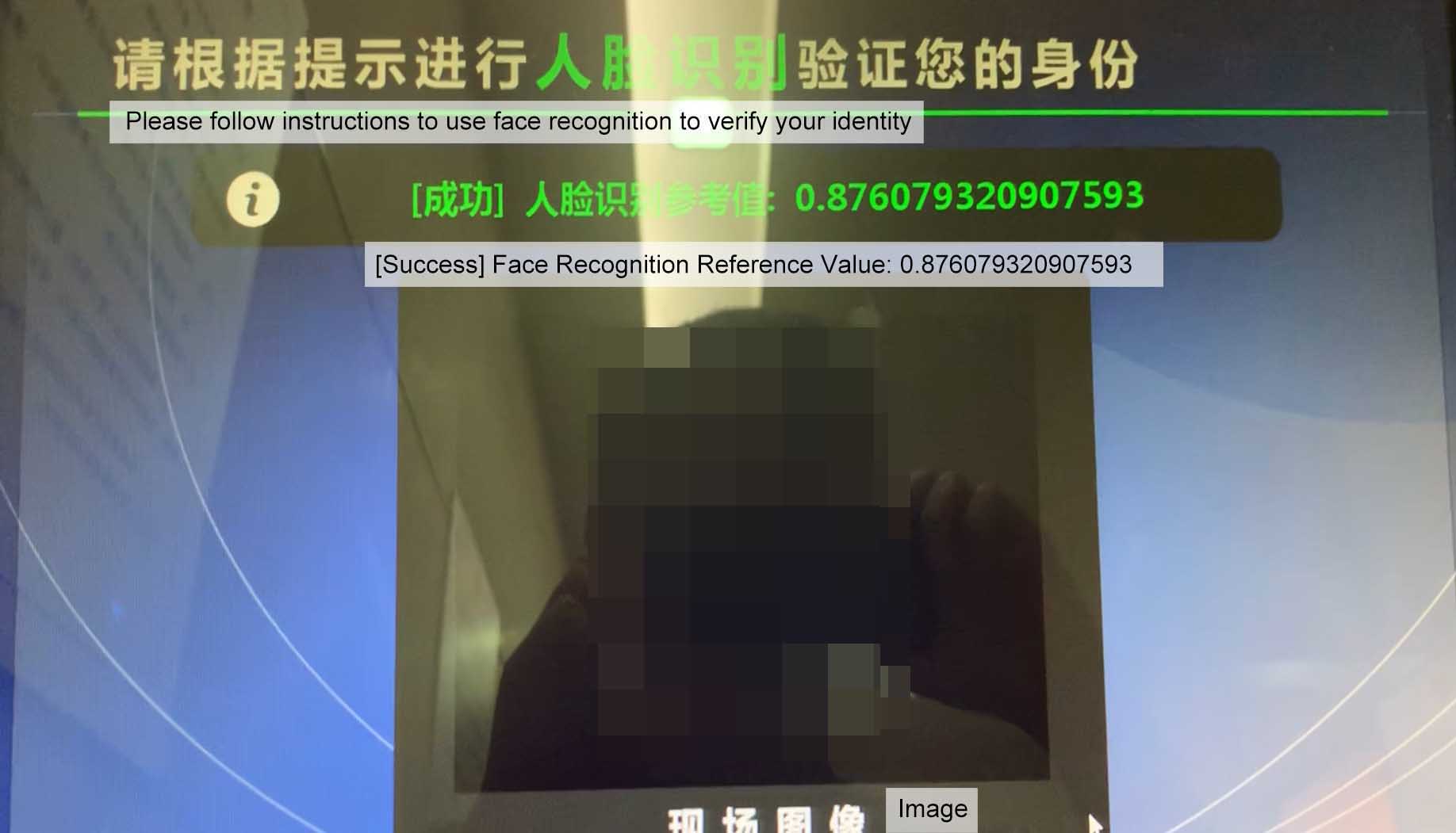}
    \caption{An self-service desk in Chinese entry \& exit bureau.}
    \label{fig:police}
\end{subfigure}\quad\quad
\begin{subfigure}{0.2\textwidth}
    \centering
    \includegraphics[width=\textwidth]{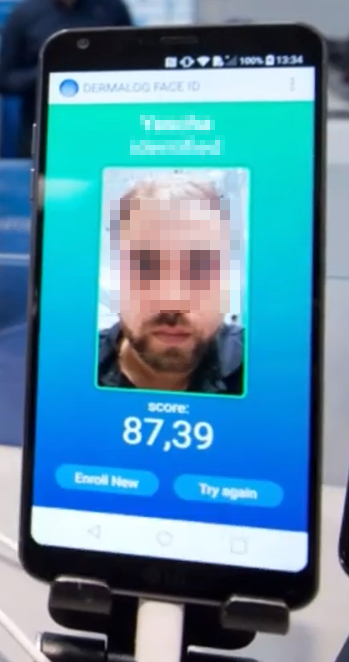}
    \caption{A Face ID app.}
    \label{fig:faceid}
\end{subfigure}
\caption{Scenarios distances are leaked.}
\end{figure*}

\subsection{Distances Leakage}

The calculated distance, from some developers' perspective, is not sensitive, according to our observation in a lot of cases. An example we encountered is a self-service machine that was deployed at the \textit{Chinese entry and exit bureau (the counterpart of immigration or boarder inspection of some countries, and also part of the police system)}. This machine authenticates users with their faces before they can touch later functions. To use this machine, a user need firstly enter his ID number and then stare at the camera for face authentication. However, the machine directly displays the similarity on the screen (see Figure.~\ref{fig:police}), which allow attackers to directly know the distance. We also found some apps with face authentication functions make similar mistakes. An app uses face as ID directly shows the matching score on the UI to users, as shown by Figure.\ref{fig:faceid}. No matter if similarity, score or confidence level displayed, they are eventually variant of embedding distance, through which attackers can acquire the distance.

To make clear the reason that developers leak distances, we discussed with five AI system developers. Unfortunately, none of them believed the distance is sensitive. What's worse is that one developer among the five believed that displaying the distance to users is necessary, because he thinks the similarity helps system managers better deploy and debug the system. Specifically, he told us that face authentication systems, especially the systems developed a couple of years ago, had poor success rate. The success rate can hardly be satisfied when firstly installed on a self service machine. The manager needs to know the real time image and confidence level to infer and handle possible factors that fails legal users' authentication, like the ambient light, irregular shadows \etc He suspects that such system deployed at complex environment must show the similarity.

\subsection{Get Embedding with Distance}
\label{sec:dist}

\zztitle{Embeddings can be recovered from distances} As we identified, if only an attacker acquires enough number of distances to a sensitive embedding, the embedding can be recovered.

\begin{figure}[htpb]
    \centering
    \includegraphics[width=0.35\textwidth]{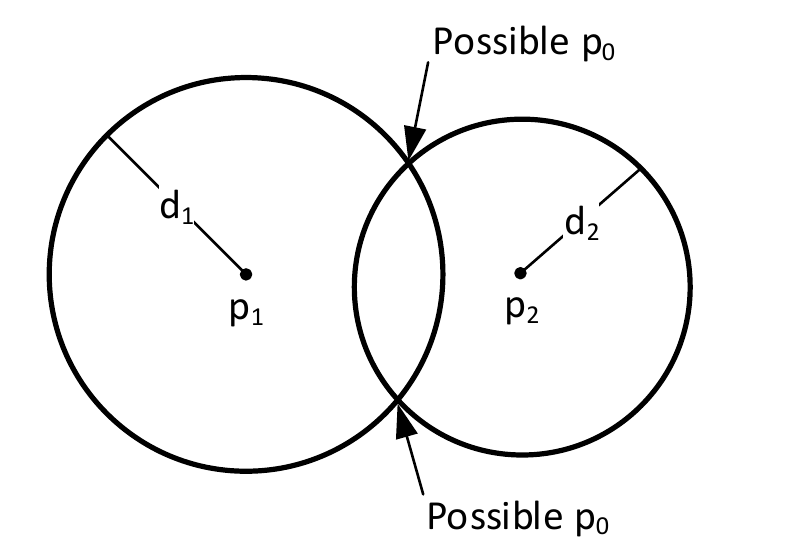}
    \caption{Illustration of two distances determine an embedding}
    \label{fig:cross}
\end{figure}

Let us consider the simplest occasion. Assuming that there is an unknown point $p_0$ (an embedding with two dimensions) in a 2D plane, if only other two points ($p_1$ and $p_2$) and their distances ($d_1$ and $d_2$) to the point $p_0$ are known, the coordinates of $p_0$ can be worked out, because $p_0$ must be located at the intersection of two circles: One centered in $p_1$ with radius $d_1$, another centered at $p_2$ with radius $d_2$. Intersection of two circles can be either two points or one point or just no intersection. Because we already know that at least $p_0$ is in the intersection, there can be either one point or two points in the intersection. As a result, if only $p_1$, $p_2$, $d_1$ and $d_2$ are known, the $p_0$ is definitely also known, with at most one false candidate, as shown by Figure.~\ref{fig:cross}

Given that a n-D embedding is used by the system, \textit{once distances to other $n$ embeddings are known, the embedding is definitely leaked}. As for the reason, embeddings are points in high dimensional space. The only difference between embeddings and 2D points is that embeddings are of higher dimension. Later we show how the attacker can work out an unknown embedding with $n$ distances and corresponding embeddings.

\zztitle{Solve equation to recover embeddings} Every time an attacker knows a distance to the embedding, she knows that the embedding is on a (n-1)-sphere. When she accumulated some different distances, she knows that the embedding is on the intersection of them. From the perspective of algebra, she gets an equation about the embedding for each (n-1)-sphere and the embedding is just the root of the equation system (shown by equation system~.\ref{eq:dist_overall}), where $\vec{x}$ is the unknown embedding, $\vec{x_i}$ and $d_i$ being the known embeddings and their distances to $\vec{x}$.

\begin{align}
|| \vec{x} - \vec{x_1} || &= d_1^2 \nonumber \\
|| \vec{x} - \vec{x_2} || &= d_2^2 \nonumber \\
&\vdots \label{eq:dist_overall} \\
|| \vec{x} - \vec{x_n} || &= d_n^2  \nonumber
\end{align}

Embedding systems usually use either cosine distance or Euclidean (L2) distance. Specifically, for embedding systems that are evaluated with $L2$ distance, the equation set the attacker needs to solve turns to equation.~\ref{eq:dist_l2}. The only obstacle to a success recovery is that how many roots the system has. If equation.~\ref{eq:dist_l2} has and only has few (like one or two) roots like the $n=2$ (circle) case, the embedding can be recovered by directly solving the equation. However, if the equation has like $O(poly(n))$ roots, the attacker still need more hard work, because she does not know which one is the embedding among these roots.

\begin{equation}
    \vec{x}^\intercal \cdot \vec{x} +A \cdot \vec{x} +D = 0 \label{eq:dist_l2}
\end{equation}

where \begin{align*}
    A&=-2\cdot \{\vec{x_1}, \vec{x_2},..., \vec{x_n}\}^ \intercal \\
    D&= \{\vec{x_1}^\intercal \cdot \vec{x_1} - d_1^2, \vec{x_2}^\intercal \cdot \vec{x_2} - d_2^2,..., \vec{x_n}^\intercal \cdot \vec{x_n} - d_n^2 \}^ \intercal
\end{align*}


As we will later show, equation.~\ref{eq:dist_l2} has at most 2 different roots and at least one root, indicating that attackers can readily recover the embedding.

To know how many different possible embeddings attackers will have with $n$ distances, we firstly check how many roots equation.~\ref{eq:proof_1} has. Because it is an ordinary linear equation system, it has only one root (shown by equation.~\ref{eq:proof_2}) given that all $\vec{x_i}$ are linearly independent (normally, two embeddings won't be parallel) and $z$ is a newly introduced constant that is independent of $\vec{x}$. If the root of equation~\ref{eq:proof_1} wants to be also the root of equation.~\ref{eq:dist_l2}, it needs also satisfy equation.~\ref{eq:proof_3}, as equation.~\ref{eq:proof_1} becomes to equation.~\ref{eq:dist_l2} if only $z$ is replaced by $\vec{x}^\intercal \cdot \vec{x}$. However, such $z$ has only up to two roots, as equation.~\ref{eq:proof_3} is an ordinary quadratic equation about $z$ which is a scalar. As a result, $\vec{x}$ can also have up to two roots, as shown by equation.~\ref{eq:root_l2}.

\begin{equation}
    z + A \cdot \vec{x} +D = 0 \label{eq:proof_1}
\end{equation}

\begin{equation}
    \vec{x} = -A^{-1}\cdot(D+z) \label{eq:proof_2}
\end{equation}

\begin{align}
    z &= \vec{x}^\intercal \cdot \vec{x} \label{eq:proof_3} \\
    &= D^\intercal BD +z\cdot\vec{1}^\intercal BD +z\cdot D^\intercal B\cdot\vec{1} +z^2\cdot\vec{1}^\intercal B\cdot\vec{1} \nonumber
\end{align}
where
\begin{align*}
    B&=(A^{-1})^\intercal \cdot A^{-1} \\
    \vec{1} &= \underbrace{\{1,1,...,1 \}^\intercal }_\text{n 1s}
\end{align*}

\begin{equation}
    \vec{x} = -A^{-1}\cdot(D+\frac{-b \pm \sqrt{b^2-4ac}}{2a}) \label{eq:root_l2}
\end{equation}
where
\begin{align*}
    a &= \vec{1}^\intercal B\cdot\vec{1} \\
    b &= \vec{1}^\intercal BD + \cdot D^\intercal B\cdot\vec{1} -1 \\
    c &= D^\intercal BD
\end{align*}

Specifically for embedding systems with cosine distance as the metric, things are similar. Equation.~\ref{eq:root_cos} tells an attacker how to calculate the embedding when she has $n$ different distances, because the embedding must satisfy equation~.\ref{eq:dist_cos}. In this case, the norm of the embedding gets lost, but the norm of embedding in cosine embedding systems is useless because two parallel embeddings will be regarded as exactly the same. One can directly normalize the embedding.

\begin{align}
    A \cdot \frac{\vec{x}}{|\vec{x}|} &= D \label{eq:dist_cos} \\
    \frac{\vec{x}}{|\vec{x}|} &= A^{-1}\cdot D \label{eq:root_cos}
\end{align}

where \begin{align*}
    A&= \{\frac{\vec{x_1}}{|\vec{x_1}|}, \frac{\vec{x_2}}{|\vec{x_2}|},..., \frac{\vec{x_n}}{|\vec{x_n}|} \}^ \intercal \\
    D&= \{ 1- d_1, 1 - d_2, 1 - d_n \}^\intercal
\end{align*}

In a word, when an attacker has accumulated $n$ distances to an embedding, she can always recover the embedding, no matter if the embedding is in Euclidean system or cosine system.

\subsection{Dimension Reduction}
\label{sec:compress}
We consider how attackers can recover the embedding when less than $n$ distances can be acquired. As we noticed, the dimension size $n$ can be huge for some embedding schemes, which makes the attack too long to succeed. Nonetheless, the entropy a piece of embedding actually contains is far less than that a $n$ dimension vector can contain. Therefore, the attacker may possibly recover the embedding even less than $n$ distances collected.

Firstly, we make clear why the actual information inside a piece of embedding is far less than the volume of the vector. We carefully reviewed the facenet embedding scheme~\cite{schroff2015facenet} and found that the number of dimensions does not influence much on the accuracy. Specifically, the 128 dimension embedding scheme has only 1 percent advantage over the 64 dimension scheme. Besides, larger embeddings including 256 and 512 dimensions both results in worse accuracy, indicating that a 64 dimension vector can already well accommodate all the information the neural network extracts. Another reason resulting in the information sparsity is that the use of dropout~\cite{JMLR:v15:srivastava14a} in CNN training. To avoid over-fitting, developers intentionally and randomly shut off some neurons during a training iteration, which encourages the network to behave similarly in different neurons. As a result, the outputs of different neurons would be similar. Therefore, the output of a network layer has a lot of redundancy.

\begin{figure}
\begin{subfigure}{0.22\textwidth}
    \centering
    \includegraphics[width=\textwidth]{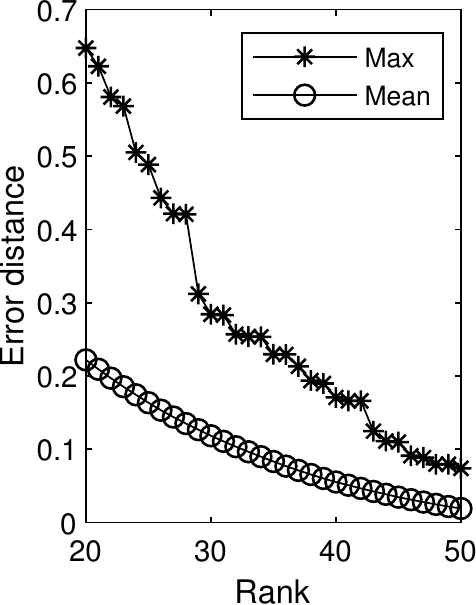}
    \caption{Error distances versus ranks of PCA on facenet-128 dimension embeddings.}
    \label{fig:svd}
\end{subfigure}\quad \begin{subfigure}{0.22\textwidth}
    \centering
    \includegraphics[width=\textwidth]{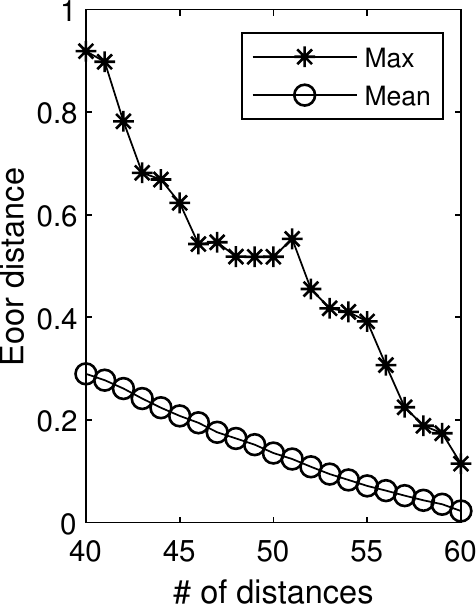}
    \caption{Error distances versus the number of distances on facenet-128 dimension embeddings.}
    \label{fig:compress}
\end{subfigure}
\caption{Error distance for PCA and for recovery method based on SVD}
\label{fig:svd_two}
\end{figure}

Second, we investigate how much entropy a piece of embedding actually contains. We calculated facenet-128 embeddings for 400 randomly selected images from the LFW dataset. Then we launched PCA over these embeddings with SVD. The 400 embeddings were regarded as row vectors and put into a matrix ($M$) row by row for SVD and PCA. Figure.\ref{fig:svd} shows the distance between embeddings in $M$ and their PCAed ones in $\tilde{M}$ at different ranks. As we can see, when the rank reaches 33, the distance goes below 0.1 which is totally negligible, considering that the threshold for differentiating people is often set to over 1.2. In other words, we can use a 33 dimensional vector to represent a 128 dimensional embedding nearly losslessly.

Third, we upgrade the method in Section~\ref{sec:dist} to allow it work with less distance observations, with the help of SVD and our finding that embeddings are of low entropy. Given that an attacker has $m$ distance observations instead of $n$ that is the number of dimensions of an embedding, the attacker can still construct the equation.~\ref{eq:dist_l2}. However, she cannot solve the equation because the equation obviously has infinite roots. The equation helps the attacker to know that the embedding is in a high dimensional manifold but cannot help her get the exact coordinate.

Considering that embeddings are of low entropy, we assume that rank $m$ is enough for representing any embedding. Therefore, a $n$ dimension embedding $e$ has negligible distance from its PCAed version, \ie, we can regard $e = e_m\cdot\Sigma_m\cdot V_m^\intercal + \delta$ in which $e_m$ is $m$ dimensional, $\Sigma_m$ being $m$-by-$m$, $V_m$ being $n$-by-$m$ and $|\delta|$ far less than $|e|$. Therefore, equation.~\ref{eq:dist_l2} turns to equation.~\ref{eq:dist_l2_svd}, given that $\vec{x}=(\vec{\tilde{x}}^\intercal\cdot\Sigma_m\cdot V_m^\intercal)^\intercal + \delta$, where $\vec{\tilde{x}}$ is a $m$ dimensional column vector and $\Delta$ being the result of $\delta$ which is negligible ($|\Delta| \ll |D|$) because $\delta$ is negligible.

\begin{equation}
    \vec{\tilde{x}}^\intercal\cdot\Sigma_m\cdot V_m^\intercal \cdot V_m\cdot\Sigma_m\cdot\vec{\tilde{x}}  +A \cdot V_m\cdot\Sigma_m\cdot\vec{\tilde{x}} +D + \Delta= 0 \label{eq:dist_l2_svd}
\end{equation}

Surprisingly, equation.~\ref{eq:dist_l2_svd} now can only have up to two roots, considering that there are only $m$ variables ($\vec{\tilde{x}}$) but the rank of any matrix does not exceed $m$ ( $V_m^\intercal \cdot V_m$ is a $m$-by-$m$ identity matrix and $A \cdot V_m$ also $m$-by-$m$). The method to solve equation.~\ref{eq:dist_l2_svd} is very similar to equation.~\ref{eq:dist_l2} and the root of equation.~\ref{eq:dist_l2_svd} is shown by equation.\ref{eq:root_l2_svd} if we neglect $\Delta$. As you may find, $\vec{x}$ here has exactly the same form with equation.~\ref{eq:root_l2} (No more SVD matrices involved), with $\vec{x}=(\vec{\tilde{x}}^\intercal\cdot\Sigma_m\cdot V_m^\intercal)^\intercal$ put in. The only difference between the root of equation.~\ref{eq:dist_l2} and that of $\vec{x}$ in equation.~\ref{eq:dist_l2_svd} is that $A^{-1}$ now refers to the pseudo inverse of $A$.

\begin{equation}
    \vec{\tilde{x}} = -\Sigma_m^{-1}\cdot V_m^{-1} \cdot A^{-1}\cdot(D+\frac{-b \pm \sqrt{b^2-4ac}}{2a}) \label{eq:root_l2_svd}
\end{equation}

This suggests the attacker that she can use equation.~\ref{eq:root_l2} to work out the embedding with negligible error directly even when not enough distances collected, if only the number of distances is larger than the rank required for PCA the embeddings with satisfied quality. Besides, the more distances collected, the less the error would be.

To verify that such method indeed works, we used the method to recover 300 embeddings randomly selected from the aforementioned 400 LFW images with some of the images selected from the rest 100 to query distances. Figure.~\ref{fig:compress} shows the recovery error against the number of distances the attacker can get. As we can see, when the attacker has more than 53 distances, the recovery error can drop to below 0.1 which is negligible.

However, by comparing Figure.~\ref{fig:svd} and Figure.~\ref{fig:compress}, we find that the error distance is slightly larger than that got by PCA when the number of distances is exactly the rank of SVD. Theoretically, the recovery loss totally stems from $\delta$ that is a result of PCA error. Nonetheless, when solving equation.~\ref{eq:dist_l2_svd}, $\Delta$ is treated as zero, which makes the root of $\vec{\tilde{x}}$ slightly deviated. Therefore, the error of the recovered embedding consists of two parts: 1) PCA error which is a result of neglecting non-principal components; 2) Root deviation that is a result of neglecting the deviation made by the PCA error to the equation. As a consequence, for a given error cap, the attacker needs more distances than the number of rank.

For the higher dimension face scheme in the facenet family, the required number of distances does not increase. After having recovered embeddings with facenet-512, we found higher dimension schemes even converges faster. For the facecet-512 scheme, it costs the attacker only 39 distances to drop the average error distance below to 0.063 (10\% of the threshold, see Table.~\ref{tab:tar_model}.) We believe it is because embedding models with better accuracy in fact extract features better. Therefore less orthogonal features are needed to achieve the same accuracy. More dimensions are a result of anti-over-fitting.

\subsection{No-box assumption}
\label{sec:nb}
The method in Section.~\ref{sec:dist} does not work with no box assumption. The method uses embeddings of the query images to construct matrix $A$, which can be done by either generating by the attacker herself or querying the embedding model with white box or black box assumptions. However, with our no box assumption, the attacker even does not know what embedding model the system uses. As a result, the attacker cannot prepare matrix $A$ for the recovery method and thereby cannot launch the attack.

A distance calculated in one embedding domain may be applied to another embedding domain. Given that the authentication system uses an embedding model $\mathcal{E}$ to generate embeddings, a distance ($d$) the attacker gets thereby is calculated with the $\mathcal{E}$ embeddings of two images. Imagine in another embedding system, say $\mathcal{E}'$, the distance $d'$ of the same two images would be calculated with their $\mathcal{E}'$ embeddings. Because $\mathcal{E}$ and $\mathcal{E}'$ in fact do the same thing though are trained with different models, we believe the distances for the same given pair of images under these two embedding domain have large similarity, \ie $f(d) \approx d'$, where $f(\cdot)$ is a linear mapping to fix output range. In other words, when $\mathcal{E}$ thinks a pair of faces are similar, $\mathcal{E}'$ has high chances to agree with it, and vice-versa.

Based on this assumption, the attacker can solve the equation in another embedding space. For example, the attacker does not know what model an authentication system uses but still gets a list of distances with a list of face images. She just arbitrarily calculates the facenet-128 embeddings of the images and directly uses the acquired distances to solve equation.~\ref{eq:dist_l2}. Then she gets a facenet-128 embedding and believes it is equal to the facenet-128 embedding of the victim image.

\begin{figure}
    \centering
    \includegraphics[width=0.4\textwidth]{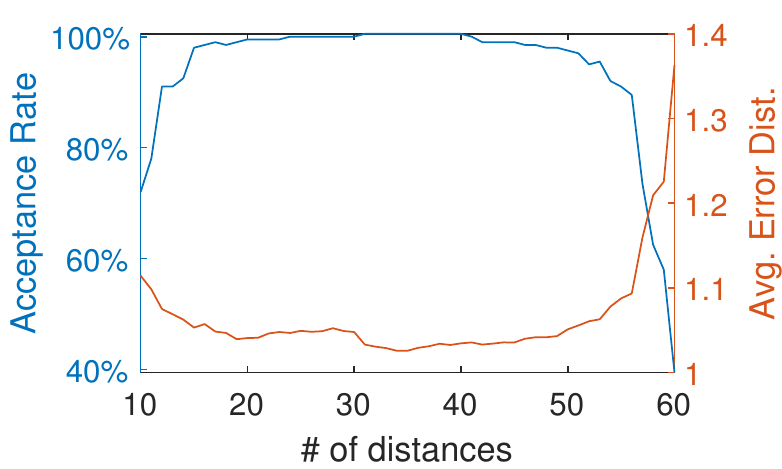}
    \caption{Average error distance and the rate of acceptance versus different number of distances the attacker uses}
    \label{fig:no_box}
\end{figure}

We also evaluated the performance of the embedding recovery with the no box assumption. We assume the authentication system uses facenet-512 model to generate embeddings and calculates distance while the attacker uses facenet-128 model to recover the embedding for 200 victim image from LFW. To be noticed is that facenet-128 and facenet-512 are totally different embedding schemes: they use L2 and cosine distance respectively and are trained on different data sets (see Table.~\ref{tab:tar_model}). In this case, the embeddings the attacker can recover are facenet-128 embeddings. We show the average distance errors from the recovered ones to the facenet-128 embeddings of the original images, and also show the ratio the recovered ones can be regarded as from the victim using a stricter threshold (1.2). Figure.~\ref{fig:no_box} shows the results.

Different from Figure.~\ref{fig:svd_two}, where the error distance decreases with increasing number of distances the attacker uses, Figure.~\ref{fig:no_box} shows an decreasing first and then increasing trend. The optimal point lies in 33, which means attackers can have least error distance with 33 distances collected for one victim. As for the reason, we believe it is because the distance in fact comes with errors resulted from the different embedding systems. The PCA scheme guarantees that more components involved and less error would be. Therefore, more distances lead to less error. However, the distance the attacker gets actually is not the distance in the attacker's embedding domain, but borrowed from the authentication system's, which is only a kind of approximation. As a result, every distance brings in information about the victim embedding but also noises simultaneously. Before the optimal, each distance cancels more SVD losses while after the optimal, the added principal components provide less distance gain than the noises, the error distance therefore increases.

With this trick, the no-box assumption turns to white box assumption. The only consequences are that 1) The embedding the attacker gets is not of the original type but the attacker's own type. 2) The corresponding image may be regarded as another image of the victim.

\section{Embedding Recovery Model}

\begin{figure*}[ht]
    \includegraphics[width=0.95\textwidth]{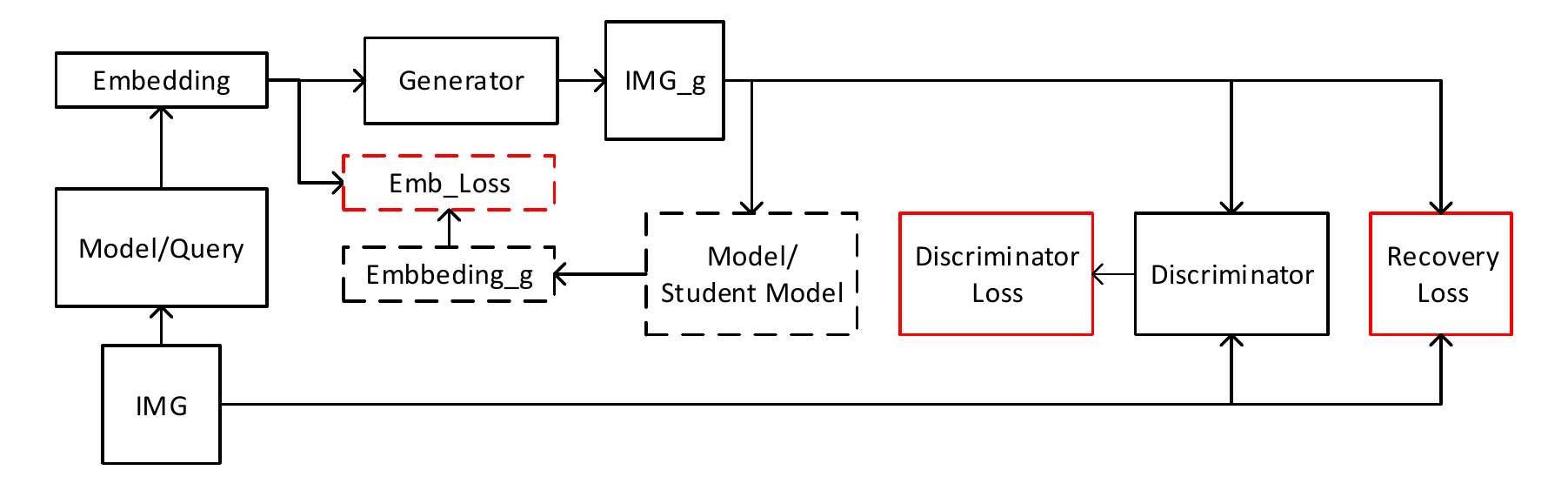}
    \caption{Overview of the GAN-like recovery framework.}
    \label{fig:gan_ov}
\end{figure*}

In this section, we introduce the recovery framework we designed. The recovery method can help attackers convert the embeddings the attacker acquired in Section.~\ref{sec:emb} to a profile image of the victim.

\subsection{Overview}
Inspired by GAN and VAE (Variational Auto Encoder), we designed the framework shown by Figure.~\ref{fig:gan_ov}. The framework mainly consists of a brand new embedding to image generator, a discriminator and several loss functions.

We assumed that the attacker owns a bunch of images to train our recovery model (denoted as $R$), which can be acquired from public face data sets. These images are firstly converted to face embeddings with the same type as the embedding to be handled (denoted as $M_t$). For black and white assumptions, $M_t$ refers to the model the authentication system uses to calculate embeddings. As the assumptions already assumes the attacker can query any image for its embedding, this step can be done straightforwardly. For no-box attackers, $M_t$ refers to the model the attacker used in Section.~\ref{sec:nb}.

\textbf{The embeddings instead of randomly generated noises are sent to the generator}, which is the main difference from ordinary GAN framework and also a key innovation of our work. The generator generates images for each input embedding. Then the generated images (IMG\_g) are sent to different modules for loss generation.

The generated images are used for generating three kinds of loss: discriminator loss, recovery loss and embedding loss. The three loss items are then used to direct updating the generator.

\subsection{Generator}
The generator does not follow classical GAN generator design. We also tried to fit classical structures to the problem setting but found none performed well.

\begin{figure}[ht]
    \includegraphics[width=0.5\textwidth]{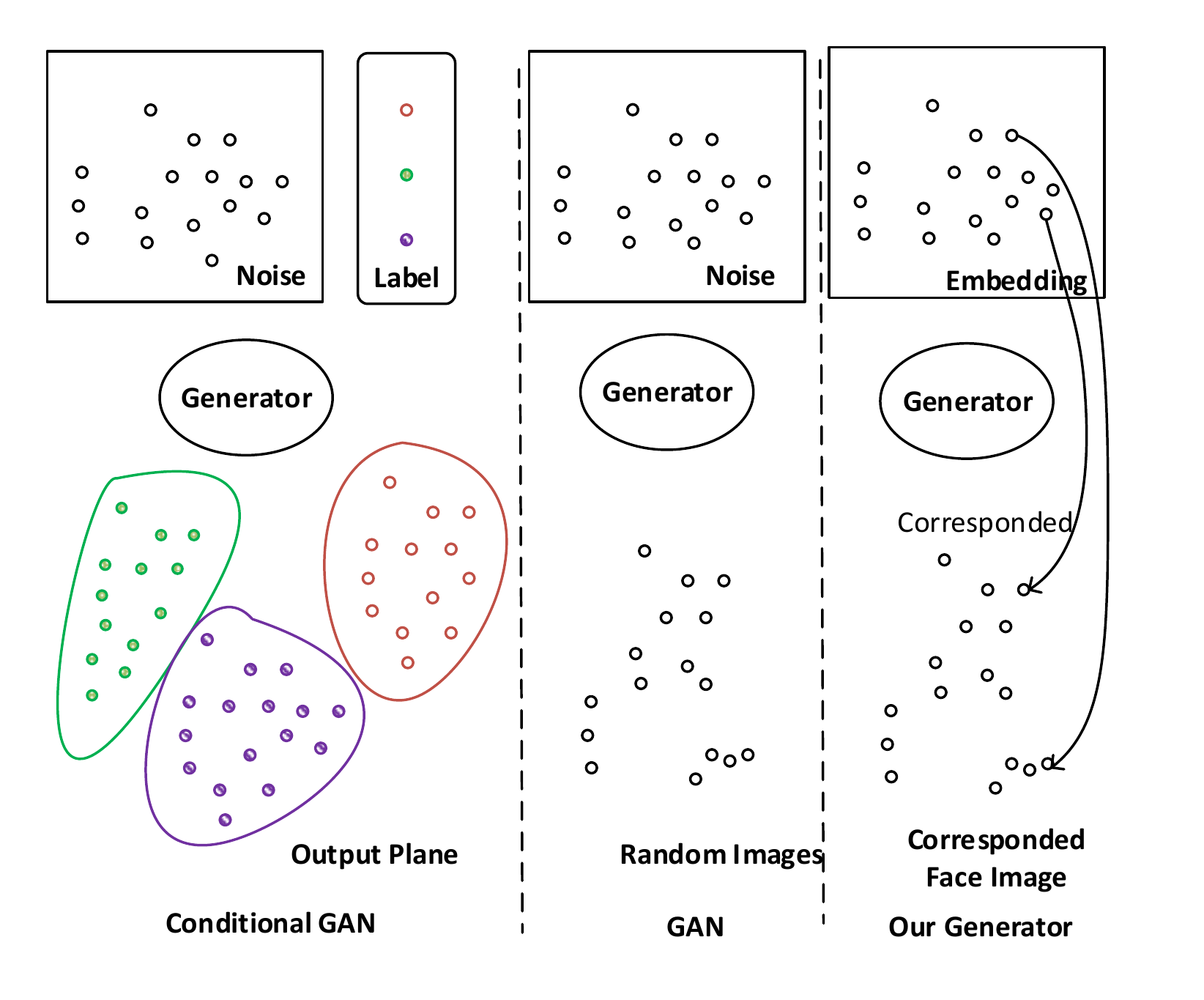}
    \caption{Comparison of Conditional GAN generator, GAN generator and our generator.}
    \label{fig:gen_comp}
\end{figure}

The generator takes embeddings as input instead of noises because in our case, we expect the model to have generalization capability over embeddings. Ordinary GANs have generalization capability over noise field, but does not limit the output, resulting in meaningful but not controlled images. However, in our setting, we need the generated images exactly corresponding to their input embeddings. Conditional GAN (cGAN) has generalization capability over noise field under the constraint of the label. If we regard embeddings as labels, cGAN indeed can make output corresponding to embeddings. However, cGAN has no generality on label, meaning that it can only generate images with seen labels, which does not satisfy our requirement. In contrast to GAN family, our generator has generalization over embeddings and also output corresponded face images. Figure.~\ref{fig:gen_comp} illustrates the difference between our generator and the generator of the other two mentioned GANs.

\begin{figure*}[ht]
    \includegraphics[width=0.95\textwidth]{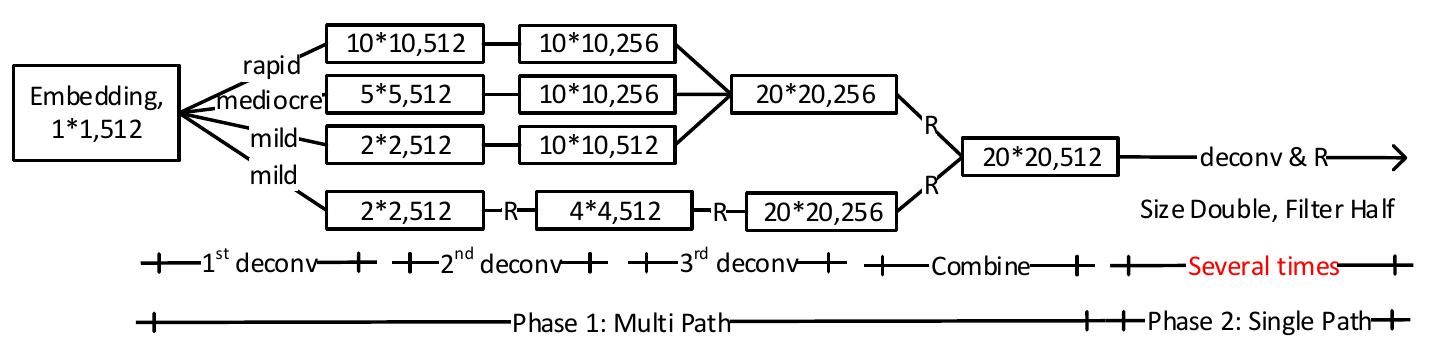}
    \caption{The design of our generator.}
    \label{fig:gen_design}
\end{figure*}

\begin{figure}[htbp]
    \includegraphics[width=0.4\textwidth]{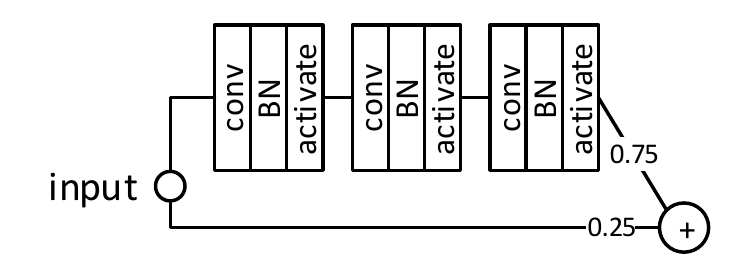}
    \caption{Residual structure used in the generator.}
    \label{fig:res}
\end{figure}

The generator we designed has two phases: multi path phase and single-path phase. Figure.~\ref{fig:gen_design} shows the design of our generator for 512 dimension embedding input. The first phase, \ie multi path phase tries to extract information from the input embedding in different paces, aiming at adapting embeddings generated by models of different types. For our 512 dimension embedding recovery case, the rapid branch directly deconvolutes the embedding from 512 dimension to 512 10*10 tiny images. In contrast, the mild branch firstly deconvolutes it into 2*2 and then 10*10. The different paces result in different grade information extraction. These branches are then combined together after they reach the same size, providing a unified processing method for the later phase. The second phase, single path phase, generates gradually clearer and larger images by concentrating channels. The data repeatedly passes deconvolution unit following by a residual convolution unit (Figure.~\ref{fig:res}). The deconvolution unit enlarges the generated images by fusing multiple channels, during which the size of the images is doubled while the channels halved, followed by the residual unit which rectifies the images without changing the image size.

\subsection{Recovery Loss}
To encourage the generator generate images corresponding to their embeddings, we add a loss item forcing the generated image (IMG\_g) similar to the original image (IMG). The loss penalizes the model according to the difference between IMG and IMG\_g. Equation.~\ref{eq:loss_r} shows the recovery loss.

\begin{equation}
    L_{r} = || IMG - IMG_g ||_1 \label{eq:loss_r}
\end{equation}

Here we use $L_1$ norm to measure the loss instead of $L_2$ because we found $L_2$ pays more attention to background comparing with $L_1$. $L_2$ cares more about larger difference while usually the background part varies more from one image to another than the face part of an image.

\subsection{Embedding Loss}
To better stimulate the generator to recover corresponded images, we also seek help from face embedding models. We deliver the generated images (IMG\_g) to a face embedding model (denoted as $M_e$) to get their embeddings (Emb\_g). We introduce the embedding loss (L\_e) to penalize the difference between the embedding of the input images and the embedding of the generated images, as shown by equation.~\ref{eq:loss_e}. The type of norm used by $L_e$ depends on the norm the face embedding model chooses, as it describes the difference between embeddings best.

\begin{equation}
    L_e = || M_e(IMG) - M_e(IMG_g) || \label{eq:loss_e}
\end{equation}

For the white box and no-box assumptions, attackers can readily use $M_t$ as $M_e$. However, an black box attacker can only query $M_t$ while the gradients that are required for training are missing there. In other words, given $x_0$ and $f(x_0)$, attackers has no way to know $f'(x_0)$ which are necessary for instructing the training optimizer. In this case, $L_e$ can be directly set to constant 0.

To also allow black box adversaries to experience the assistance from embedding loss, an attacker can either 1) train a substitutional model as $M_e$ or 2) use an open source model instead.  

For the first case, attackers can employ the teaching-student method\cite{hinton2015distilling} to construct a model from the embeddings and images such that the model has approximately the same behaviour with the $M_t$. Thus, with $x_0$ to $x_n$ and $f(x_0)$ to $f(x_n)$, the attacker can train a model $g$ such that $g(x_0)$ to $g(x_n)$ are all approximately equal to $f(x_0)$ to $f(x_n)$. Then the attacker can use $g$ to substitute the $f$ as $M_e$ to get gradients and direct the updating, expecting that $g'$ is also an approximation to $f'$.

For the second, if only a model $g$ is also a face embedding model, the model $g$ can be used to measure the quality of the generated images, though it may have utterly different characteristics from the original model $f$. The attacker can expect an open source model to help evaluate the similarity between a generated images and the original training image. Thus, they expect $|| g(x_1) - g(x_2) ||$ with some positive relationship with $|| f(x_1) - f(x_2) ||$.

Multiple $M_e$ can be used together to get better quality. Because of the over-fitting characteristics of deep learning models, it is expected that one face embedding model focuses on some kinds of features on the input. By combining multiple different models, they can learn more features to better weight the quality loss.

\subsection{Discriminator}
The discriminator guarantees that the generated images indeed look like images containing faces.

The discriminator we employed follow the standard discriminator from WGAN-GP. The loss is directly used as our discriminator loss($L_d$). WGAN-GP needs to maintain Lipschitz function to calculate the Wasserstein distance. It penalizes gradient for every independent sample. The discriminator we use also drops all batch normalization layers (BN), and after every convolutional operation, we add a small Residual Block just like our generator. In the end, the output of the discriminator will be a scalar that is the confidence the discriminator thinks the input is real. 

\subsection{Training Process}
We follow the GAN training process to train our model, \ie training generator and discriminator in turn. When training the generator, 3:1:1 to 2:1:1 was found to be the best ratio for $L_r$ , $L_d$ and $L_e$. Learning rate is decayed 0.02 for every epoch. We train the generator five times after every single discriminator training iteration.

\section{Evaluation}
We evaluate the recovery model we designed in this section. To understand how much risk the recovered embeddings result in, we show the images recovered from the recovered face embeddings.

\subsection{Target Embedding Model}
We have chosen four face embedding models as our target $M_t$. They are a self trained Inception Resnet, Clarifai online face embedding model, the current version Facenet and an old version but more well-known Facenet model.

\begin{table}[htpb]
    \centering
    \begin{tabular}{|c|c|c|c|c|} \hline
    Model & \makecell{Emb. \\ Length} & \makecell{Distance \\Type} & TH & Acc. \\ \hline
       \makecell{Residual \\ Inception Network}  & 1792 & Cosine & 0.78 & 92.1\% \\ \hline
        \makecell{Clarifai Online \\ Face Embedding~\cite{clarifai}} &  1024  & Cosine & 0.55 &98.1\%\\ \hline
        \makecell{Facenet \\ 20180402-114759}  & 512  & Cosine & 0.63 &97.6\%\tablefootnote{We used dlib as alignment tool, which resulted in lower accuracy than that shown on git.}\\ \hline
        \makecell{Facenet \\ 20170512-110547} & 128 & L2\tablefootnote{Facenet required squared L2. we use L2 instead, as they are equivalent for comparing only.} & 1.28 &97.1\%\tablefootnote{We used dlib as alignment tool, which resulted in lower accuracy than the that shown on git.}\\ \hline
    \end{tabular}
    \caption{Target models. TH is the threshold the model would regard two embeddings from the same person. Acc. is the accuracy under the threshold we set.}
    \label{tab:tar_model}
\end{table}

\zztitle{Self trained Resnet model}
To evaluate recovery quality for the case that the target authentication system uses a self-trained model, we tried to train a model with little hard work, which is what developers of small enterprises with ML demand do. We trained the model with popular network structure on an open face dataset. The structure we chose is inception-resnet-v1, the training dataset is CASIA-Webface, and the test dataset being LFW dataset. We added cross-entroy loss over the ``Additive Margin Softmax'' after densing the embedding, which turns the model to a classifier for training. Because the model was trained with dense as classifier, the embedding can be measured by cosine distance. The model got only moderate accuracy as no fancy Deep learning tricks were added, which imitates small enterprise models.

\zztitle{Online Model}
Mainly to test if our recovery model can recover images under a pure black box adversary assumption, we add to our target list a totally black box face embedding model, \ie online model. we surveyed popular online face embedding models and found Clarifai the most popular one\footnote{Clarifai ranked the first when Google searching ``Face Embedding API''.}. Clarifai provides almost all platform SDKs to allow developers directly access their server for embedding generation, during which the model is never exposed to developers.

\zztitle{Facenet}
Facenet\cite{schroff2015facenet} is the first well-known work for face embedding, as its triplet loss greatly improved face embedding performance. Facenet also is the most popular open source face embedding work\footnote{Facenet ranked the first when Google searching ``Face Embedding''.}. The most popular implementation of Facenet was found at\cite{facenet_git}\footnote{It appeared next to the paper of Facenet when searching ``Facenet'' on Google.}. According to the history of the git page, the author published two 128 dimension face embedding model and recently updated to the 512 dimension version. We evaluate on both the current 512 Dimension model and one of the previous 128 Dimension model.

\subsection{Evaluation Metric}
To confirm if the recovery model ($R$) generated images indeed look like embedding owners, we introduce a judger model $M_j$ to measure the distance between the original images and the generated images, and use the distance to quantify privacy leakage extent. Specifically, we hire Facenet-512 model as the judger $M_j$. If the distance is below the commonly used threshold (0.63 for our evaluation), we regard the recovery quality loss acceptable and say a successful attack.

Besides, we also recruited five volunteers to subjectively rate the similarities between original face images and the recovered images, in case the model generates images solely cheating $M_j$.

\subsection{Experiment Setup}
We set up a platform (Table.~\ref{tab:setup}) to run all the following evaluation experiment.
\begin{table}[htbp]
    \centering
    \begin{tabular}{c|c}
        CPU      &  Core i9-9900k\\ \hline
        Memory   &  64G   \\\hline
        GPU      &  2080Ti 11G \\\hline
        Platform & Tensorflow 1.10
    \end{tabular}
    \caption{Experiment environment for evaluation.}
    \label{tab:setup}
\end{table}

We use most of images of LFW (12000) to train the recovery model ($R$) while the left 384 images were used for testing. The time consumed to train a recovery model is around 10 - 13 hours. Before evaluating the recovery model, we firstly get the embeddings of testing images. The testing images are firstly sent to our recovery method in Section.~\ref{sec:dist} for recovery. Unsurprisingly, there is no recovery error encountered (less than eps). Then we use the recovered embeddings to evaluate the recovery model.

To be noticed is that we have up to three models involved in a single training: 1) The target model $M_t$, 2) $M_e$, the model for $L_e$ generation; 3) the judger model $M_j$ which is set to facenet-512 in our evaluation.

\begin{table}[htpb]
    \centering
    \begin{tabular}{|c|c|c|c|c|} \hline
    Target Model & \makecell{Success \\ Rate}  & \makecell{Avg. Q. \\ Loss} & \makecell{Avg. \\ Q. Rating} & \makecell{Avg. \\ S. Rating} \\ \hline
       \makecell{Wide Residual \\ Inception Network}  & 62.76\% & 0.5901  &3.9 & 3.7\\ \hline
        \makecell{Clarifai Online \\ Face Embedding} &  58.00\% \tablefootnote{Clarifai failed to return embeddings for some LFW images, so only 250 images were left for testing}  & 0.5987  &3.8 &4.0\\ \hline
        \makecell{Facenet \\ 20180402-114759}  & 92.19\%  & 0.4186  & 4.5 & 4.6 \\ \hline
        \makecell{Facenet \\ 20170512-110547} & 93.75\% & 0.4135  & 4.6 &4.6 \\  \hline
    \end{tabular}
    \caption{Black box evaluation result. Avg. Q. Loss is the average distance between a generated image and the original image. Q. Rating (Quality Rating) is the subjective rate for quality while S. Rating (Similarity Rating) is the subjective rating for similarity between generated and original.}
    \label{tab:black_res}
\end{table}

\subsection{Pure Black Box Recovery}
\label{sec:black}
We evaluated all the four $M_t$ under the black box adversary assumption. For the three models which eventually we have the details about the models, we set the embedding loss ($L_e$) to constant as if attackers do not know the model but train our recovery model.

\zztitle{Evaluation Result} As we can see from Table.~\ref{tab:black_res}, the recovery model trained for the Facenet-128 dimension model got the best overall recovery performance, while got the worst for the Clarifai model. The recovery quality for the two Facenet models are nearly perfect, though are less surprising for the other two.

Table.~\ref{tab:samples} (column 2 to 5 for black box assumption) shows some recovered samples for each model, which indicates similar conclusion. For the cases with best recovery quality, \ie Facenet 512D and 128D, the recovered images clearly show faces that can be affirmatively regarded as another version of the victim. Even for the worst case, \ie Clarifai, there are still considerable similarities between the recovered and original, indicating huge privacy threaten to embedding owners.

Subjective rating also goes with the quality metric. Our volunteers agreed that the recovered images against the two Facenet models have high fidelity and most (over 90\%) of them can be thought as real photos taken for their owners.

\newcommand{\mysize}{0.12}

\begin{table*}[t]
    \centering
    \begin{tabular}{c c c c c c c}
    Original & \makecell{Facenet \\ 128D} & \makecell{Facenet \\ 512D} &
    \makecell{Clarifai \\ Online Model} & \makecell{Self Trained \\ Model} &
    \makecell{Substitutional \\ Model Assisted} & \makecell{White Box}
    \\ \hline
       \includegraphics[width=\mysize\textwidth]{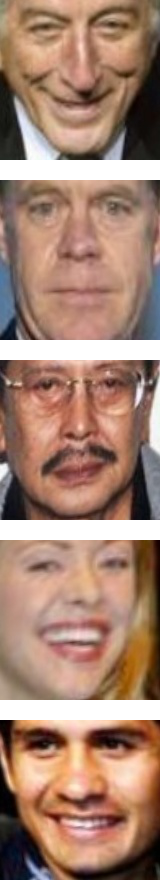} &
        \includegraphics[width=\mysize\textwidth]{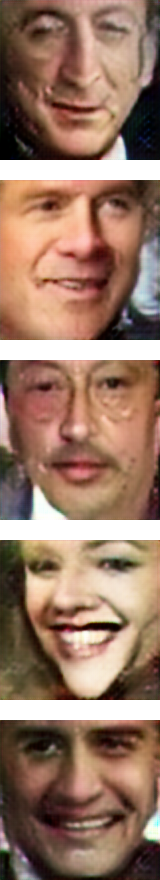} &
        \includegraphics[width=\mysize\textwidth]{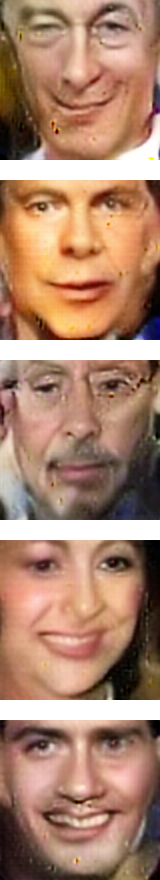} &
        \includegraphics[width=\mysize\textwidth]{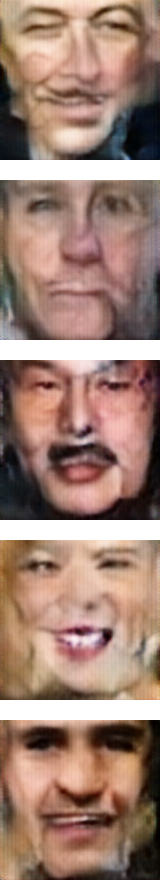} &
        \includegraphics[width=\mysize\textwidth]{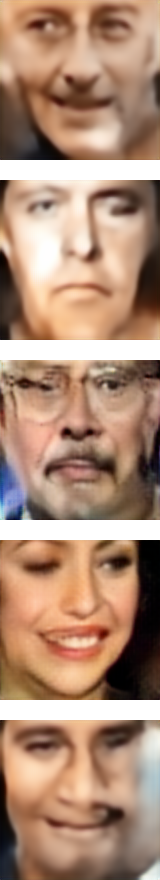} &
        \includegraphics[width=\mysize\textwidth]{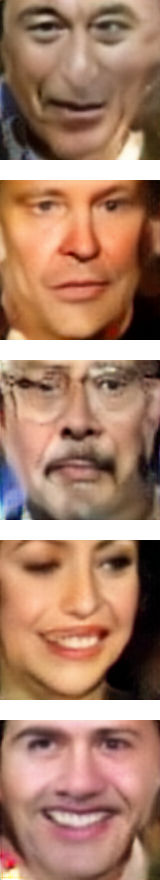} &
        \includegraphics[width=\mysize\textwidth]{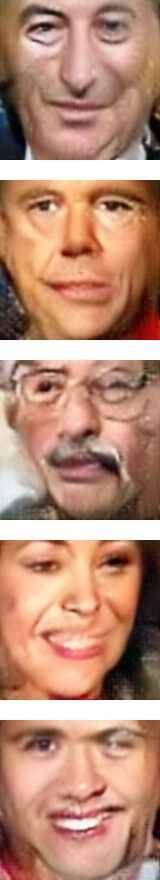}
    \end{tabular}
    \caption{Recovered samples during evaluation. }
    \label{tab:samples}
\end{table*}

\begin{figure}
    \centering
    \includegraphics[width=0.5\textwidth]{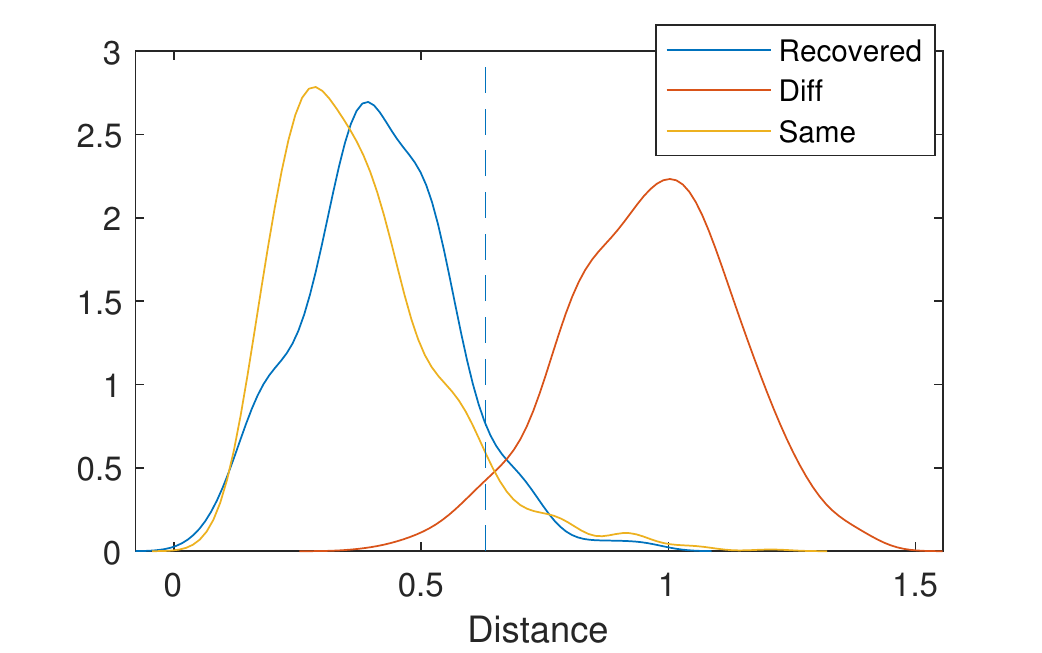}
    \caption{Probability Density Function of distances between embeddings of the same person, different people and \{original, recovered\} images respectively. The vertical line shows the threshold for the Facenet 512D model to regard as the same person.}
    \label{fig:pdf}
\end{figure}

\zztitle{Distance Distribution} A recovered image can be regarded as another photo of the embedding owner, from the perspective of similarity. Figure.~\ref{fig:pdf} shows the distribution of distances between two images from the same person, two images from different people and the pair of original and recovered images. As we can see, the threshold line can clearly tell apart the distances of the same person (left, smaller) and different people (right, larger). However, the distance between original images and recovered images has a nearly identical distribution with that of the same person, indicating that a recovered image can be regarded as from the embedding owner.

\zztitle{Recovery Performance vs embedding length} The recovery quality has nearly no correlation with embedding length, which goes with our PCA experiment conclusion.




\zztitle{Recovery Quality vs Accuracy}
The recovery performance neither has strong correlation with embedding model accuracy. It is believed that better recognition accuracy is a result of richer information delivered by embeddings. Similarly, the richer information can result in better recovery quality. The results showed that this is not always true. The recovery for online Clarifai model has relatively poorer quality while this model has very high recognition accuracy. Except the Clarifai case, the rest three models approximately follow this rule. Facenet models have both better recognition accuracy and recovery quality than the self trained resnet model.

We still believe the recovery quality depends on target embedding model's accuracy. We analyzed Clarifai online model and found that, though the model yield a very high recognition accuracy, the distance between two embeddings from different people is still pretty low. Specifically, distance from different people and distance from the same person have close absolute value, though they have clear boundary so they can be told apart from each other. We believe this is the reason of relatively lower quality for the high accuracy model. Besides, we have no idea if Clarifai model was trained with LFW. We are sure that the other three models have never seen our data to train the recovery model, \ie LFW. However we have no any knowledge about Clarifai model's training process. So, it is possible that LFW was also used as part of their training data, which later interfered our training process, as training data are always over-fitting points of the model (it will later be discussed in another scenario).

From another perspective, we believe the difficulties for recovering embeddings of different models are different. Embedding models' capabilities to describe mappings are different, some are of higher while some are of lower. Usually, embedding model with higher describing capability results in higher recognition accuracy. While our recovery model only has limited capability to describe mapping. If the target model is so strong (higher accuracy) such that the recover model does not have enough capability to describe its inverse mapping, the recovery quality would be low. In this case, attacker need employ a stronger generator to deal with the model.

\subsection{Recovering Training Data}
\label{sec:gen_train}
Usually people do not care prediction quality on training data. However we still evaluated the recovery quality on training data of the $M_t$, because we found there are chances that ML developers devote later collected user data to fine tune their model. As a result, user data which originally were test data become training data.

We used 1472 images from CASIA-WEB database (the training data of Facenet 512 model) and tested its recovery quality on the recovery model ($R$) for the Facenet 512 model we got in section. \ref{sec:black}. Surprisingly, the success rate 88.18\% is even lower than that on unseen test data (92.19\%).

We conjecture this is because the training data is too over-fitting for $M_t$. Specifically, for $M_t$, the distribution of training data's embedding and that of unseen data's embedding actually are slightly different. However, our recovery model is trained with data that are never seen by $M_t$, resulting in misleading and inappropriate inverse mapping description on $M_t$'s training data. As a result, our recovery model is not that suitable to recover $M_t$'s training data.

\subsection{Substitutional Model Assisted Black Box }
\label{sec:sub_model}
We simulated a black box adversary who uses an open source model as $M_e$ to assist her recovery model training.

We take the latest Facenet implementation, \ie Facenet 512D,  as $M_t$. When training the recovery model, we use the old version Facenet 128D model as $M_e$.

\begin{table}[htbp]
    \centering
    \begin{tabular}{|c|c|c|c|c|} \hline
       \makecell{Evaluation \\ Method} & Suc. Rate & \makecell{Suc. \\ Inc.} & Q. Loss &  Q. Inc. \\ \hline
         \makecell{Substitutional \\Model Assisted} & 94.53\%  &  2.54\% & 0.4149 & 0.89\% \\ \hline
         \makecell{Training \\Data Recovery} & 88.18\% & & 0.4727 & \\ \hline
         \makecell{White Box} & 91.41\% & & 0.4340 & \\ \hline
    \end{tabular}
    \caption{Evaluation results of substitutional model assisted black box scenario, training data recovery and white box scenario.}
    \label{tab:sub-black}
\end{table}

As we can see from Table.~\ref{tab:sub-black}, the success rate and quality both increased, with the assistance of an old version model .

We believe the most important factor for the improvement is model diversity. With a different model supervising the image generation, it is more possible that the generator can generate images with less defects, given that every embedding model may neglect some important feature of the input image which may be taken care by another model. It is possible that if even more models are added to supervise the generator training (by adding multiple $L_e$ of different model) process, the generator can perform better. However we also noticed that GPU memory may not be able to accommodate more models. We cannot even successfully put a 128D Facenet model to assist our generator for the 1024D and 1792D model, because of GPU memory limitation. We believe GPUs with more memory may help attackers achieve multiple model assisted training.

\subsection{White Box Recovery}

We also evaluated recovery quality under the white box assumption. We trained the recovery model for Facenet 512D with $L_e$ constructed also by Facenet 512D.

Unfortunately, we ended up with a even worse results than black box model, as shown by Table.~\ref{tab:sub-black}. It goes contrary with our common sense. Usually we believe a model works better under white box model because more information is given. The only explanation we can find is that recovery model prone to over-fitting if it is also supervised by the target model ($M_t$ as $M_e$). This explanation is coincide to the performance improvement in Section.~\ref{sec:sub_model} and Section.\ref{sec:gen_train}. The generator already exploited the mapping from embedding to image, while the opinion of another model as embedding loss actually corrects some points neglected by the target model. However, the neglected part would be impressed if the same model again interfere the mapping learning.

Considering the observation, a white box attacker would better hire another model to assist recovery model training. 
\section{Related Works}
Papernot \etal surveyed \cite{papernot2018sok} security and privacy issues around ML from nearly all aspects. However, our work cannot be categorized according to the survey. According to their survey, privacy issues during inference mainly result in three attacks: membership inference, training data extraction and model extraction. Apparently, our work cannot be simply put into any among the three. We list some very relevant works for further reference. 

\subsection{Training Data Security}

Fredrikson \etal showed in \cite{fredrikson2015model} that part of training data can be reconstructed given confidence values revealed along with predictions. Specifically for their face recognition study, they reconstructed meaningful images of a victim who is in the training set. Besides, given a blurred image of a training image, an attacker can identify the victim from the training set or determine that the images owner is not in the training set. The work is an extension from their previous work \cite{fredrikson2014privacy} which proposed the model inversion attack. Generically, Reza \etal proposed a quantitative investigation on information leakage by machine learning models~\cite{shokri2017membership}. They specifically focus on membership inference problem which determines if a record is in the training set of a given model. Nasr \etal proposed a method to protect membership privacy using adversarial regularization~\cite{nasr2018machine}.

Under the framework of differential privacy, Abadi \etal developed new techniques to train model with differential privacy on training data \cite{abadi2016deep}

To protect training data when multiple parties should jointly learn a model, Reza \etal proposed a kind of collaborative learning framework that enable them to learn an accurate model without sharing training data~\cite{shokri2015privacy}. However, Hitaj \etal showed that even in collaborative learning scenarios, attackers can still recover training data at some extent, mainly by utilizing GAN.

To be noticed is that our work is working on another track comparing with these works, though we all targets recovering data related to ML models. Training data can be recovered from models mainly because ML models usually have very large gradient or special featured gradient resulted from over-fitting at training points. While our work focuses on recovering unseen data in which these characteristics are missing.

\subsection{Model Security}
Tramer \etal proposed \cite{tramer2016stealing} a method to extract ML models in the ML-as-a-service scenario, indicating that models can be readily stolen by attackers if only queries is allowed toward a model. Apart from models, hyper-parameters are also investigated and found likely to be stolen~\cite{DBLP:conf/sp/WangG18}. Oh \etal showed that even optimization procedure can be inferred~\cite{oh2017towards}.

Besides through querying, researchers proposed to infer the model structure through running traces, Naghibijouybari \etal showed that performance counters collected during GPU running can be used to infer model hyper-parameters~\cite{naghibijouybari2018rendered}. Besides performance counter, cache is another source for model inference~\cite{yan2018cache,hong2018security}.

Hanzlik \etal developed MLCapsule, aiming at preventing model stealing or reverse engineering while keeping user input locally stored\cite{hanzlik2018mlcapsule}.

\section{Discussion}
In this section, we discuss some potential countermeasures to embedding leakages and also some possible future work directions.
\subsection{Defense}
\zztitle{Better Developer Education}
ML library and SDK documentations should clearly tell developers that distances can only be exposed to authorized managers and can never be displayed to normal users. Developers should also learn case studies about embedding leakages so they will not leak distances inadvertently.

\zztitle{One-way Model} Just like one-way hash function, ML developers may design models in a style that the reverse mapping of a model cannot be easily worked out by attackers. Hash functions employ computation subroutines that are hardly to reverse.  However, basic units used by ML models today, like pooling, convolution, activation, are all partially or totally reversible, resulting in reversible models. We hope ML developers could design model structures that cannot be easily reversed.

Ideas from protecting models, like \cite{hanzlik2018mlcapsule}, can also be borrowed to make the model irreversible. 

\zztitle{Introducing Noises}
Inspired by VAE, we propose adding to embeddings noises, such that the accuracy of using embeddings does not decrease sharply while the image recovery turns to be of low quality. 

VAE usually produces images that are obviously dimmer, which is because of distributions rather than determined numbers are produced by the encoder. As a result, to tolerate the variances (noises), the decoder produces dimmer images comparing with GAN.

We suspect an embedding with noises may similarly results in lower recovery quality. Therefore, ML developers may intentionally add noises to embeddings before comparing images. As a result, leaked embeddings can only help attackers produce vague images.
\subsection{Future Work}

\zztitle{Better GAN} Our work employs a GAN like structure to generate images. The recovery quality can be largely improved if more advanced techniques in GAN can be borrowed. The borrowed discriminator is still far from perfect as for some models the images generated still come with defects and damage. ML professionals may re-design the structure so the recovery quality can be improved.

\zztitle{Quantifying information on embedding} There are differential privacy works targeting quantitatively measuring information leakages on a published model. It's also valuable to construct a framework to evaluate the carried information on distances, with which we can know the upper bound of potential leakage related to embeddings.
\section{Conclusion}
The face similarity, which looks like insensitive, is displayed to users in some scenarios. Our work demonstrate that face similarity actually contains rich information about the user. Once dozens of such similarities acquired by an attacker, she can readily recover the embedding of the victim face, with our proposed embedding recovery equations. What's worse is that the embedding is equally sensitive as the victim face. To support this point, we designed a GAN like recovery model that converts the recovered embedding back to face images. The recovery quality is good as most of the recovered images can be regarded as from the victim from the perspective of the judger model. We call the community to pay attention to the unobserved leakage and pushes developers to avoid such leakages in the future.



\bibliographystyle{ACM-Reference-Format}
\bibliography{main}

\end{document}